\documentclass{acm_proc_article-sp}

\usepackage{url}
\usepackage{algorithmic}

\begin{document}

\title{A Multi-Stage CUDA Kernel for Floyd-Warshall}

\numberofauthors{2}
\author{
\alignauthor Ben Lund\\
	\affaddr{University of Cincinnati}\\
	\affaddr{Department Of Computer Science}\\
	\affaddr{814 Rhodes Hall}\\
	\affaddr{Cincinnati, OH 45221}\\
	\email{lund.ben@gmail.com}\\
\alignauthor Justin W. Smith\\
	\affaddr{University of Cincinnati}\\
	\affaddr{Department Of Computer Science}\\
	\affaddr{814 Rhodes Hall}\\
	\affaddr{Cincinnati, OH 45221}\\
	\email{smith5jw@mail.uc.edu}
}
\maketitle

\begin{abstract}
We present a new implementation of the Floyd-Warshall All-Pairs Shortest Paths algorithm on CUDA.\footnote{After
completion of this paper, the authors discovered that similar results were obtained by Bulu\c{c}, Gilbert and Budak\cite{BGB09}.}
Our algorithm runs approximately 5 times faster than the previously best reported algorithm.
In order to achieve this speedup, we applied a new technique to reduce usage of on-chip shared memory and allow the CUDA scheduler to more effectively hide instruction latency.
\end{abstract}




\section{Introduction}
\label{sec:Intro}
Graphics Processing Units are parallel processors offering high FLOPS/sec for low cost.
NVIDIA's CUDA framework makes it practical to take advantage of this resource for high performance computing by providing a scalable programming model and a stable and comprehensible machine model to program against.
Even with the tools provided by CUDA, making full use of the inherent processing power of a GPU can be a challenge, since the speed of a computation may not be limited by the speed of the processing units, but can also be limited by memory bus bandwidth or problems with thread scheduling.

Algorithms that find the length of the shortest path between every pair of vertices in a weighted and directed graph solve the all pairs shortest paths problem.
This class of algorithms, including the Floyd-Warshall algorithm (e.g. \cite{DEW06}), have a wide variety of applications in areas as diverse as bioinformatics, routing, and network analysis.
The Floyd-Warshall algorithm is a standard approach for all pairs shortest paths problem that works with negative edge weights, doesn't suffer performance degradation for dense graphs, and has predictable execution regardless of the underlying data.
It has can be decomposed into $n^3$ atomic tasks and runs in $\Theta(n^3)$ time, where $n$ is the number of vertices in the graph.

Implementations of the Floyd-Warshall algorithm for CUDA have been presented in the past.
Harish and Narayanan \cite{HN07} presented an implementation assigning a single thread for each atomic task.
This execution time of this approach was limited by the time necessary to pass data over the bus between global memory and the multiprocessors.
Katz and Kider \cite{KK08} achieved a speedup of $5-6\times$ over Harish and Narayanan's simple implementation by using a blocked approach that performs several tasks for each data element passed over the global memory bus.
It requires three 32 by 32 tiles of data elements from the adjacency matrix to be stored in on chip shared memory for the duration of the execution of a thread block, which limits the flexibility of the GPU thread scheduler, resulting in significant exposed instruction latency.

Our implementation achieves a further speedup of over $5\times$ over the implementation of Katz and Kider.
This includes a $2.1-2.3\times$ speedup from reducing the instruction count and using less expensive instructions, and a $2.3-2.5\times$ speedup from reducing the shared memory required by each thread block, allowing the thread scheduler to more effectively hide latency.
Our improvements to shared memory usage consist of making efficient use of the on chip registers and staging the algorithm so that not all shared data dependencies for a block need to be in shared memory at any one time.
On the \emph{NVIDIA Tesla C1060}, our implementation is able to 
solve APSP for any graph with single precision edge weights containing 16,384 vertices in 53.06
seconds.\footnote{From this, one can see the implied constant (in seconds) 
for the run time of this $\Theta(n^3)$ algorithm is around $1.2 \cdot 10^{-11}$.}
This is over 150x as fast as a basic Floyd-Warshall implementation running on our CPU.

This paper is organized as follows.
In Section \ref{sec:PreviousWork}, we discuss previous work in optimizing the Floyd-Warshall algorithm.
In Section \ref{sec:Sequential}, we discuss the blocked Floyd-Warshall algorithm used by Katz and Kider (\cite{KK08}) in more depth.
We discuss the improvements we made to the basic blocked Floyd-Warshall in Section \ref{sec:LatencyOptimizedCUDA}.
In Section \ref{sec:Results}, we present our results and show that they are close to the best possible on our hardware.

\section{Previous Work}
\label{sec:PreviousWork}

An important improvement in cache utilization for the Floyd-Warshall algorithm
was found by Venkataraman et al.\ \cite{VSM03}. 
(Ferencz and Diament presented a similar improvement a few years earlier in \cite{DF99}.)
The method of Venkataraman et al.\ first partitions the incidence matrix into multiple tiles.
From this, they noted that each tile of the matrix may be processed more extensively 
(i.e., one may iterate the ``outer loop'' multiple times over the tile)
before proceeding to another tile.  This approach allows for a significant improvement in cache utilization.
They reported a speedup of between $1.6\times$ and $1.9\times$ over the standard implementation on 
their targeted architecture.

Further work on improving cache utilization was done by Penner and Prasanna \cite{PP06},
and also Han et al.\ \cite{HFP06}.
The work of Han et al.\ is particularly useful. 
They created a program that generates an implementation of the Floyd-Warshall algorithm 
that is optimized for a specific architecture, including the use of SIMD vectorization 
that is available on modern PC processors.

Bondhugula et al.\ \cite{BDFWS06} consider the efficient use of FPGAs (Field Programmable Gate Arrays) 
to improve performance of Floyd-Warshall.
They noted (published in 2006) that their result was 
``an improvement by a factor of tens over the CPU implementation''.

Harish and Narayanan demonstrated the use of CUDA to solve APSP (along with several other graph algorithms) \cite{HN07}.
Soon after, a significant improvement was made by Katz and Kider \cite{KK08}.
Their implementation utilized the blocking algorithm described by Venkataraman et al. \cite{VSM03}.
Their approach resulted in a $5\times$-$6.5\times$ improvement over that of Harish and Narayanan.

\section{Blocked Floyd-Warshall Algorithm on CUDA}
\label{sec:Sequential}
\subsection{The Floyd-Warshall Algorithm}
\begin{figure}
\begin{algorithmic}[1]
\STATE \COMMENT{Assume all $w_{xy}$ are initialized to the weight of edge $(x, y)$}
\FORALL{$1 \leqslant k \leqslant n$}
	\FORALL{$1 \leqslant i \leqslant n$}
		\FORALL{$1 \leqslant j \leqslant n$}
			\STATE $w_{ij} \gets min( w_{ij}, w_{ik} + w_{kj})$
		\ENDFOR
	\ENDFOR
\ENDFOR
\end{algorithmic}
\caption{The Floyd-Warshall Algorithm}
\label{FW-Alg}
\end{figure}
The Floyd-Warshall algorithm is a familiar algorithm to solve the all pairs shortest paths problem.
Let $G=(V, E)$ be a weighted directed graph, where $V = \{ v_1, v_2, \ldots, v_n \}$.
Let $w_{ij}$ be the weight assigned to the edge $(v_i, v_j)$.
If $(v_i, v_j) \notin E$, then $w_{i,j} = \infty$, and for all $v_i \in V$, $w_{i,i}=0$. 
Otherwise, $w_{i,j} \in \mathbb{R}$, such that no negative cycles exist.
Let $d^{(k)}_{i,j}$ be the shortest path from $v_i$ to $v_j$ that (only) passes through some subset of
$\{ v_1, v_2, \ldots, v_k \}$. 
The basic Floyd-Warshall algorithm successively calculates $d^{(k)}_{i,j}$ by iterating over $k$ and updating $w_{i,j}$ to be the shortest path from $i$ to $j$ using only vertices $0$ through $k$. (see Figure \ref{FW-Alg})
At completion, the incidence matrix $W$ contains $d^{(n)}_{i,j}$ for all i and j.

Harish and Narayanan \cite{HN07} implemented Floyd-Warshall on CUDA in the simplest and most direct way possible.
They created a kernel containing a single iteration of the innermost loop, and repeatedly assigned as many threads running this kernel as possible.
Since the GPU is designed to efficiently schedule a large number of threads, this approach works reasonably well and typically offers some speedup over the same algorithm on a CPU, but it suffers from global memory bus saturation.

It is easy to see that the core algorithm requires $3$ words to be sent from global memory to the multiprocessors for the calculation and $1$ word to be sent back to global for storage.
This is 16 bytes sent across the global memory bus for each task performed.
On our setup, the measured speed for device-to-device memcpy operations for the  \emph{NVIDIA Tesla C1060} is 77 GB/sec.
Based on this, we can calculate that we can expect to perform fewer than $4.8*10^9$ tasks per second, and in fact our measured 
speed for this algorithm is approximately $2.6*10^9$ tasks/sec (see section \ref{sec:Results}).
On the other hand, the advertised computing speed for the  \emph{NVIDIA Tesla C1060} is 933 GFLOPs/sec.
Even unoptimized, the basic calculation of finding the correct index to process, adding, and taking a minimum requires fewer than the 359 FLOPs the \emph{Tesla} is capable of performing in the measured time for a single task.

\subsection{Blocked Floyd-Warshall}
Katz and Kider \cite{KK08} recognized this problem and applied existing research (e.g. \cite{VSM03}) on improving cache efficiency of the Floyd-Warshall algorithm to make practical use of the shared memory associated with each multiprocessor in the GPU (see \cite{NVPG23} and \cite{NVPTX14} for details on the CUDA machine and programing models).
The algorithm they used is a blocked Floyd-Warshall that performs several tasks for each data element transferred across the memory bus.
 Figure \ref{BFW-Alg} shows a basic sequential blocked Floyd-Warshall algorithm.

In the blocked algorithm, the matrix is partitioned into tiles and the tasks are performed in stages.
Katz and Kider use $32 \times 32$ tiles, and $32$ tasks are performed for each data element in the matrix during each stage of the algorithm.
Each stage requires the execution of three kernels.
The first kernel calculates $32$ tasks for each data element in a single tile on the diagonal of the adjacency matrix, termed the "independent block."
Tasks in the independent block depend only on other tasks in the independent block, or on tasks that were computed in a previous stage.
The second kernel calculates $32$ tasks for each data element in each tile aligned with the independent block in either the i- or j- direction.
These are the "singly dependent blocks."
Tasks in the singly dependent blocks have one data dependency within the block, and one dependency in the already calculated independent block.
There are $\Theta(n)$ singly dependent blocks in each stage.
Finally, the third kernel calculates $32$ tasks for each of the remaining tiles.
These are the "doubly dependent blocks," and each of these blocks depends entirely on two singly dependent blocks.
Since there are no dependencies internal to a doubly dependent block, these tasks may be performed in any order.
There are $\Theta(n^2)$ doubly dependent blocks in each stage, and it is the efficiency with which this stage is performed that determines the speed of the algorithm.

In Katz and Kider's implementation of the doubly dependent kernel for this algorithm, two singly dependent tiles containing dependencies and the doubly dependent tile being processed are loaded into shared memory for each thread block.
The thread block then performs 32 tasks for each data element in the doubly dependent tile before copying the doubly dependent tile back to global memory.
Clearly, the total data sent through the global memory bus is reduced by a factor of 32 in comparison with the implementation of Harish and Narayanan.
Performing a similar bandwidth calculation as before, we would expect that, if the speed of the algorithm were limited by the memory bus, it would be capable of performing $154*10^9$ tasks/second on the \emph{NVIDIA Tesla C1060} in the ideal case, or perhaps half that if the performance were comparable to the previous algorithm.
In fact, our results (see section \ref{sec:Results}) indicate that this algorithm can perform $14.9*10^9$ tasks/second, which indicates that it is limited by the rate it performs the tasks on the multiprocessors and not by the rate it can send data between the multiprocessors and global memory.
This is not a surprise, since, in the ideal case, we would need to be able to perform a task with just 6 FLOPs before bandwidth would be the determining factor.

\begin{figure}
\begin{algorithmic}[1]
\STATE\COMMENT {Assume all $w_{xy}$ are initialized to the weight of edge $(x, y)$}
\FORALL{$0\leqslant b < n/s$}
\STATE\COMMENT{Process the independent block first.}
	\FORALL{$b*s \leqslant k < (b+1)*s$}
		\FORALL{$b*s \leqslant i < (b+1)*s$}
			\FORALL{$b*s \leqslant j < (b+1)*s$}
 				\STATE $w_{ij} \gets min( w_{ij}, w_{ik} + w_{kj})$
			\ENDFOR
		\ENDFOR
	\ENDFOR
	\STATE\COMMENT{Tasks in a singly dependent block depend on the independent block.}
	\STATE\COMMENT{i-aligned singly dependent blocks}
	\FORALL{$0 \leqslant ib \leqslant n/s$}
		\FORALL{$b*s \leqslant k < (b+1)*s$}
			\FORALL{$b*s \leqslant i < (b+1)*s$}
				\FORALL{$ib*s \leqslant j < (ib + 1)*s$}
			 		\STATE $w_{ij} \gets min( w_{ij}, w_{ik} + w_{kj})$
				\ENDFOR
			\ENDFOR
		\ENDFOR
	\ENDFOR
	\STATE\COMMENT{j-aligned singly dependent blocks }
	\FORALL{$0 \leqslant jb \leqslant n/s$}
		\FORALL{$b*s \leqslant k < (b+1)*s$}
			\FORALL{$jb*s \leqslant i < (jb+1)*s$}
				\FORALL{$b*s \leqslant j < (b + 1)*s$}
			 		\STATE $w_{ij} \gets min( w_{ij}, w_{ik} + w_{kj})$
				\ENDFOR
			\ENDFOR
		\ENDFOR
	\ENDFOR
	\STATE\COMMENT{Most blocks are doubly dependent. Notice that k is now innermost.}
	\FORALL{$0 \leqslant ib \leqslant n/s$}
		\FORALL{$0 \leqslant jb \leqslant n/s$}
			\FORALL{$jb*s \leqslant i < (jb+1)*s$}
				\FORALL{$ib*s \leqslant j < (ib + 1)*s$}
					\FORALL{$block*size \leqslant k < (block+1)*size$}
						 \STATE $w_{ij} \gets min( w_{ij}, w_{ik} + w_{kj})$
					\ENDFOR
				\ENDFOR
			\ENDFOR
		\ENDFOR
	\ENDFOR	

\ENDFOR
\end{algorithmic}
\caption{Blocked Floyd-Warshall Algorithm}
\label{BFW-Alg}
\end{figure}

\subsection{Thread Blocks and Multiprocessors in the Blocked Floyd-Warshall}
Each multiprocessor on a device of compute capability 1.3 like the \emph{NVIDIA Tesla C1060} can manage up to 1024 threads and has 16384 bytes of shared memory and 16384 one-word registers.
A program will assign a large number of thread blocks, and multiple thread blocks may be processed simultaneously on each multiprocessor, subject to the limitations just mentioned.

The blocked Floyd-Warshall algorithm described above uses a tile size of 32.
Each thread block processing a doubly dependent block copies three tiles to shared memory - the doubly dependent tile to process and two singly dependent tiles with the dependencies for doubly dependent tile.
The total shared memory consumed per block is 3 tiles  $*32^2$ words per tile $*4$ bytes per word $ + 32$  bytes for parameters $ = 12320$ bytes.
Since this is more than half of the total shared memory available per multiprocessor, only one block can be assigned to a multiprocessor at any one time.

196 threads assigned to a single multiprocessor can hide latency from register dependencies, and 512 threads is enough to hide latency of global memory access for most applications \cite{NVBPG23}.
This assumes that a significant portion of the threads to be scheduled are available when needed.
If a thread is waiting at a synchronization point, it can't be scheduled.

In this application, the kernel first copies part of the tiles it needs from global memory, and then waits at a synchronization point.
Since only one block can be assigned to each multiprocessor, when the last thread to copy data from global memory executes its instruction, there are no threads available to schedule.
Accessing global memory has significant latency (hundreds of cycles) which will be exposed when the scheduler is starved for threads.

\section{Staged Blocked Floyd-Warshall}
\label{sec:LatencyOptimizedCUDA}
We applied two rounds of improvements to a blocked implementation matching the performance reported in Katz and Kider.
First, we performed standard optimizations to reduce the instruction count and use less expensive instructions.
Primarily, this was using bit shifts instead of division or modulus operators where possible and unrolling loops.
These basic optimizations yielded a speedup of $2.1-2.3 \times$

Once we had a well optimized basic blocked implementation, we applied two techniques to reduce the shared memory consumed by each thread block.
First, we made more efficient use of the multiprocessor registers.
Second, we staged the data load so that only a fraction of the singly dependent tiles containing the calculated dependencies for the current doubly dependent tile being processed were in shared memory at any one time.
The net effect of these optimizations was to reduce the shared memory used by a thread block by a factor of nearly 12 without changing the amount of data flowing between global memory and the multiprocessors.
Enabling multiple thread blocks to be assigned to each multiprocessor enables the thread scheduler to effectively hide the latency of accessing global memory and allows more tasks to be assigned to each thread which reduces the computational overhead of calculating initial indexes.
We changed from using 256 threads per thread block to 64.
These changes resulted in an additional $2.3-2.4 \times$ speedup, for a total improvement of approximately $5.2 \times$.

\subsection{Efficient Use of Registers}
One critical observation is that each multiprocessor has more memory available in registers than in shared memory.
For a device of compute capability 1.3, there are 16384 4-byte registers and a total of 16384 bytes of shared memory.
By reading data that does not need to be shared among threads in a thread block directly to registers, the total shared memory consumed can be reduced.
This optimization will be useful when thread occupancy or the number of blocks assigned to a multiprocessor is limited by shared memory and additional registers are still available.

In this case, all of the tasks that involve updating a specific data element in the doubly dependent matrix are assigned to a single thread, and no thread needs to access the data elements assigned to another thread.
As a result, the doubly dependent tile can be read directly into registers.
Instead of reading the doubly dependent tile into an array of size $t*t$ in shared memory, each thread reads its assigned elements into a local array of size $t*t/ h$ where h is the number of threads per thread block.
In general, the compiler will create an array with constant indexes in registers.
An array that doesn't have size determined at compile time will be created in local memory, which is vastly slower.

\begin{figure*}
\centering
\includegraphics[width=3in, height=3.4in]{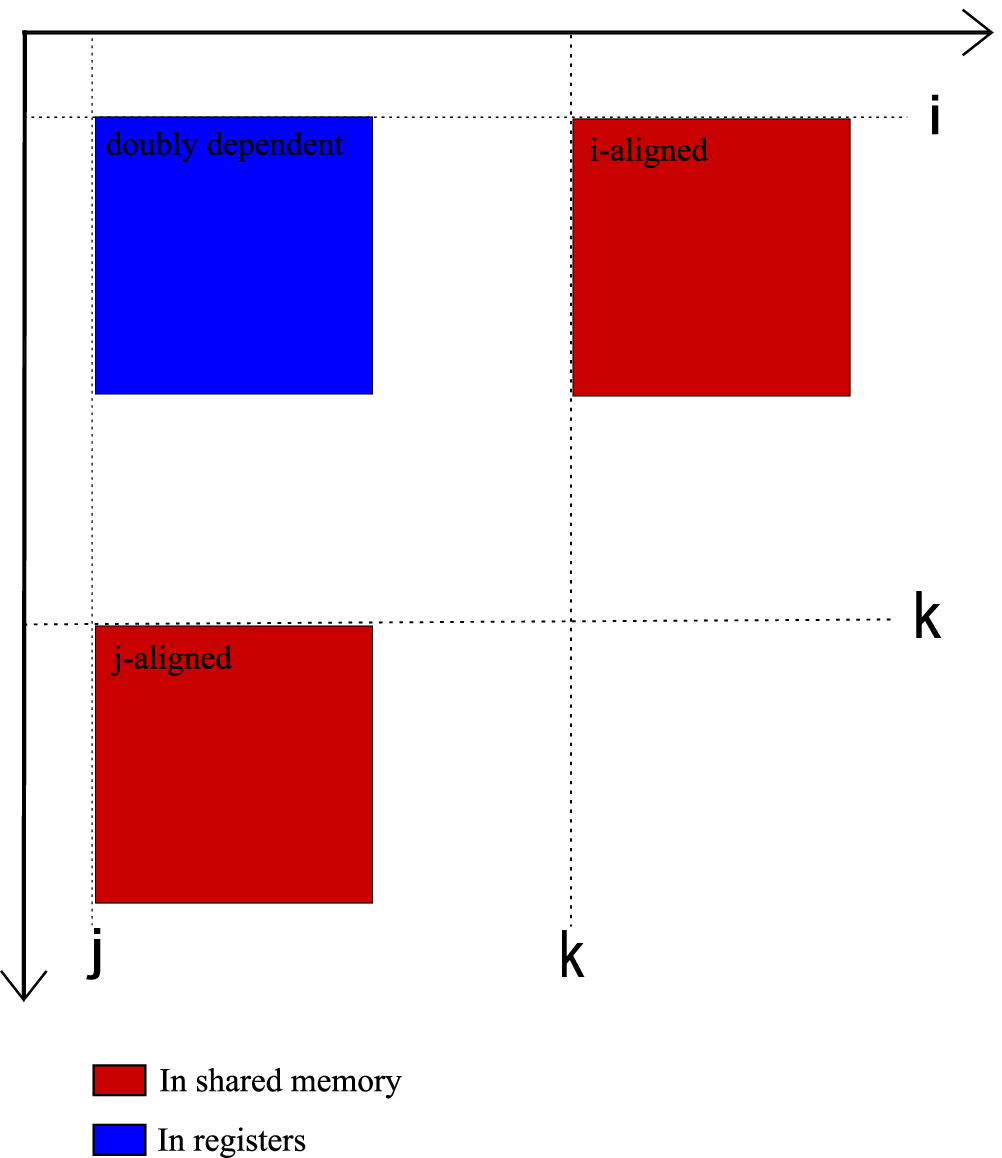}
\caption{Shared Memory Usage with Doubly Dependent Tile In Registers}
\label{RegisterMemoryUsage}
\end{figure*}

Figure~\ref{RegisterMemoryUsage} shows the memory layout in shared memory when the doubly dependent tile is stored in the registers.
This optimization reduces the shared memory consumed per block to $2*32^2+32 = 8224$, which is more than half of the available 16384.
It is still only possible to assign a single thread block to each multiprocessor, so this improvement does not by itself affect performance.

\subsection{Staged Load of Singly Dependent Tiles}
To further reduce shared memory usage, we read the singly dependent blocks in stages.
All of the tasks in the doubly dependent block in a certain $k$-range will depend a specific subset of the data elements in the singly dependent tiles.
Specifically, tasks in the range $(i,j)_k$ to $(i,j)_{k+m}$ will depend on data elements in the $i$-aligned tile in the range $(i,k)$ to $(i,k+m)$ and in the $j$-aligned matrix in the range $(k,j)$ to $(k+m,j)$.
Since the data elements of the doubly dependent tile are evenly assigned to the threads in a thread block, each thread will have the same number of tasks in a given $k$-range.
As a result, we can break the algorithm into stages, each separated by a synchronization point.
In each stage, the data dependencies for the tasks in the next $k$-range of size $m$ are loaded into shared memory arrays, the threads are synchronized, and the next $m$ tasks for each data element are performed.
Using this approach, only $t*m$ data elements for each singly dependent tile need to be loaded into shared memory at any one time.

\begin{figure*}
\centering
\includegraphics[width=3in, height=3.4in]{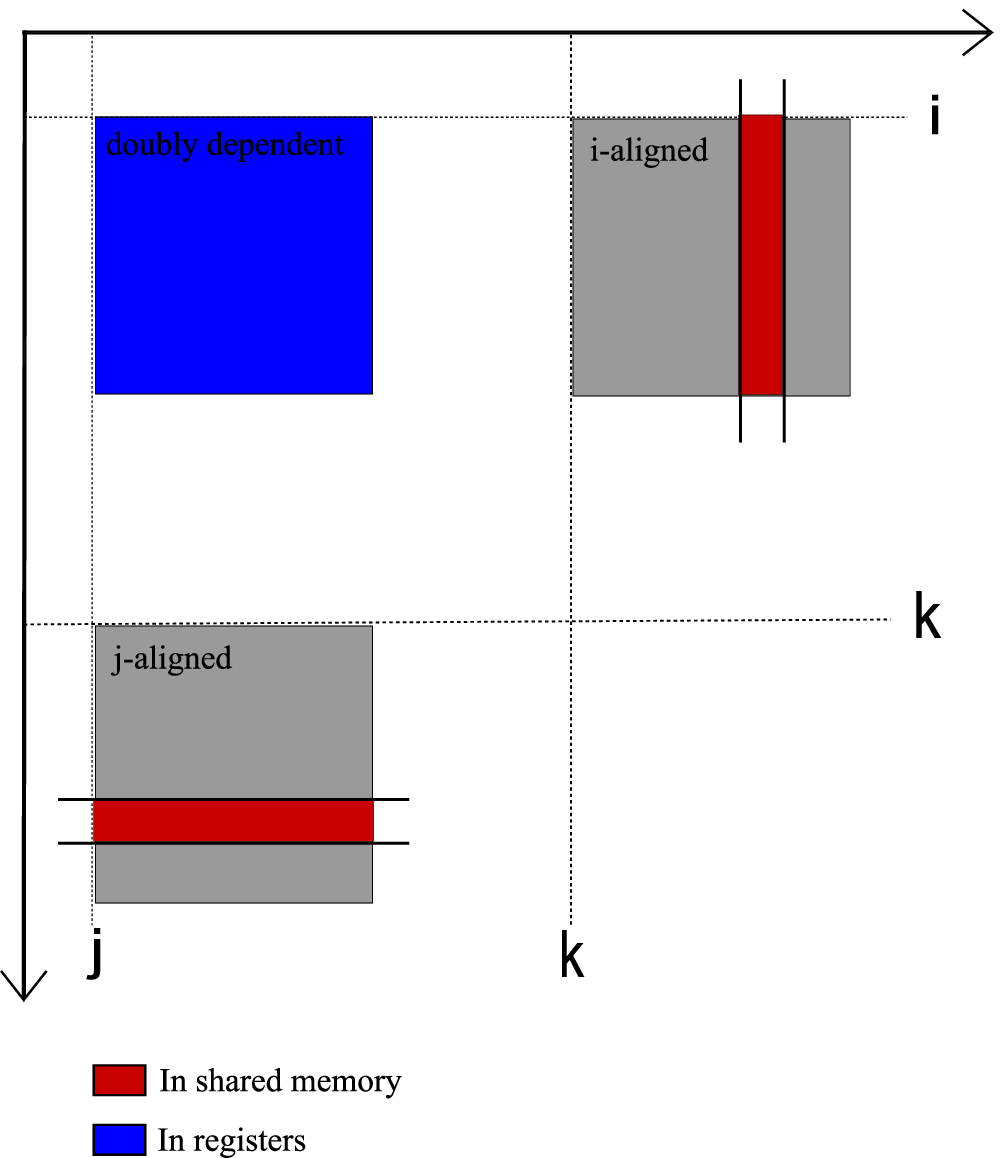}
\caption{Shared Memory Usage with Doubly Dependent Tile Updated in Stages}
\label{StagedMemoryUsage}
\end{figure*}

Figure~\ref{StagedMemoryUsage} shows the usage of shared memory during a single stage of the update of the doubly dependent tile.
In this case, the red areas of the i-aligned and j-aligned tiles contain all the dependencies for tasks in the doubly dependent tile over some range of k.
When this range is updated, the next slice of each singly dependent tile will be loaded into shared memory, and the doubly dependent tile will be updated over the next k-range.

Our specific implementation uses a tile size of 32 and stages the algorithm over 4 iterations.
The total shared memory used in this implementation is $2*32*4*4 + 32 = 1056$, so as many as 15 blocks could be run on each multiprocessor given the shared memory usage.
The limiting factors are now the total threads assigned to a multiprocessor and the registers used for each thread block.

\subsection{Implementation}
There are two significant hurdles to implementing the staged load of the singly dependent tiles.
First, it is necessary to retain coalesced access to global memory.
A single half warp that reads from global memory should access aligned and contiguous words.
Using a row-major data structure, a slice of the j-aligned tile will be composed of contiguous 16-word blocks but a slice of the i-aligned tile will not.

\begin{figure}
\centering
\includegraphics[width=2.6in, height=4.8in]{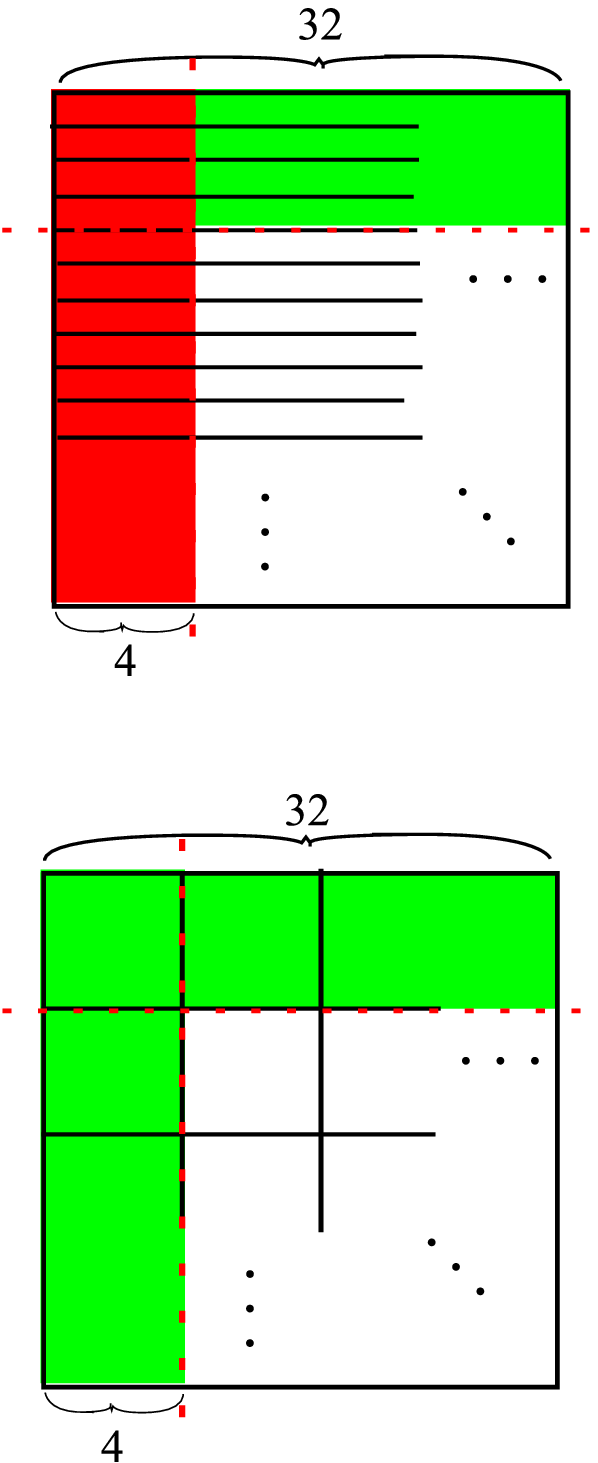}
\caption{Patterns of Global Memory Access for Row-Major and Tiled Data Orders}
\label{JustifyTiling}
\end{figure}

Figure~\ref{JustifyTiling} shows this situation.
A set of 4 adjacent columns from a row major matrix involves accessing only 4 adjacent data elements.
By tiling the data in 4 by 4 blocks, we can take 4 rows or 4 columns with 16 or more adjacent elements in either direction without increasing the load on the global memory bus.

In order to be able to read both a set of 4 columns and a set of 4 rows in contiguous 16-word blocks, we use a doubly tiled, row-major data order similar to that used by Han, Franchetti, and P\"{u}schel \cite{HFP06}.
In this ordering, each 32 by 32 tile and each 4 by 4 tile is contiguous in memory.
Within the entire matrix, the 32 by 32 tiles are arranged in row-major order, and within each 32 by 32 tile the 4 by 4 tiles are arranged in row-major order.

Second, it is necessary to avoid shared memory bank conflicts when accessing the singly dependent tiles.
Shared memory is arranged into 16 banks.
There are two patterns of memory access that allow a read from shared memory to be completed in a single cycle.
Either all of the threads in a half warp must read the same data element, or each thread must read from a different memory bank.
With all of the data elements assigned to a thread in a half warp adjacent in row-major order as they are in the Katz-Kider implementation, the shared memory access is very naturally good.
In the j-aligned tile, each thread accesses adjacent memory locations, and in the i-aligned tile, each thread in the half warp accesses the same memory location.
When the data is arranged into 4 by 4 tiles, however, this simple memory access pattern breaks down.
Threads 0, 4, 8, and 12 all access the same data element in the j-aligned tile and threads 0, 1, 2, and 3 access the same element in the i-aligned tile, resulting in 4-way data conflicts.
This data access pattern causes each shared memory access to take 4 processor cycles.
Since each task involves two reads from shared memory, this is a significant hit to performance.

To ameliorate this problem, it is important to remember that, although all of the tasks in a doubly dependent block must be performed, the order in which they are performed doesn't matter.
As a result, different threads in a single half warp can process the tasks for their data elements in any order; they are not restricted to performing them in the most natural order, with $k$ proceeding from 0 to t.
Instead, the first task performed by a thread in a given halfwarp is the task identified by the sum of the $i$ and $j$ indexes within the tile modulo 4.
This results in conflict free shared memory access.

\begin{figure*}
\centering
\includegraphics[width=4.5in, height=5.5in]{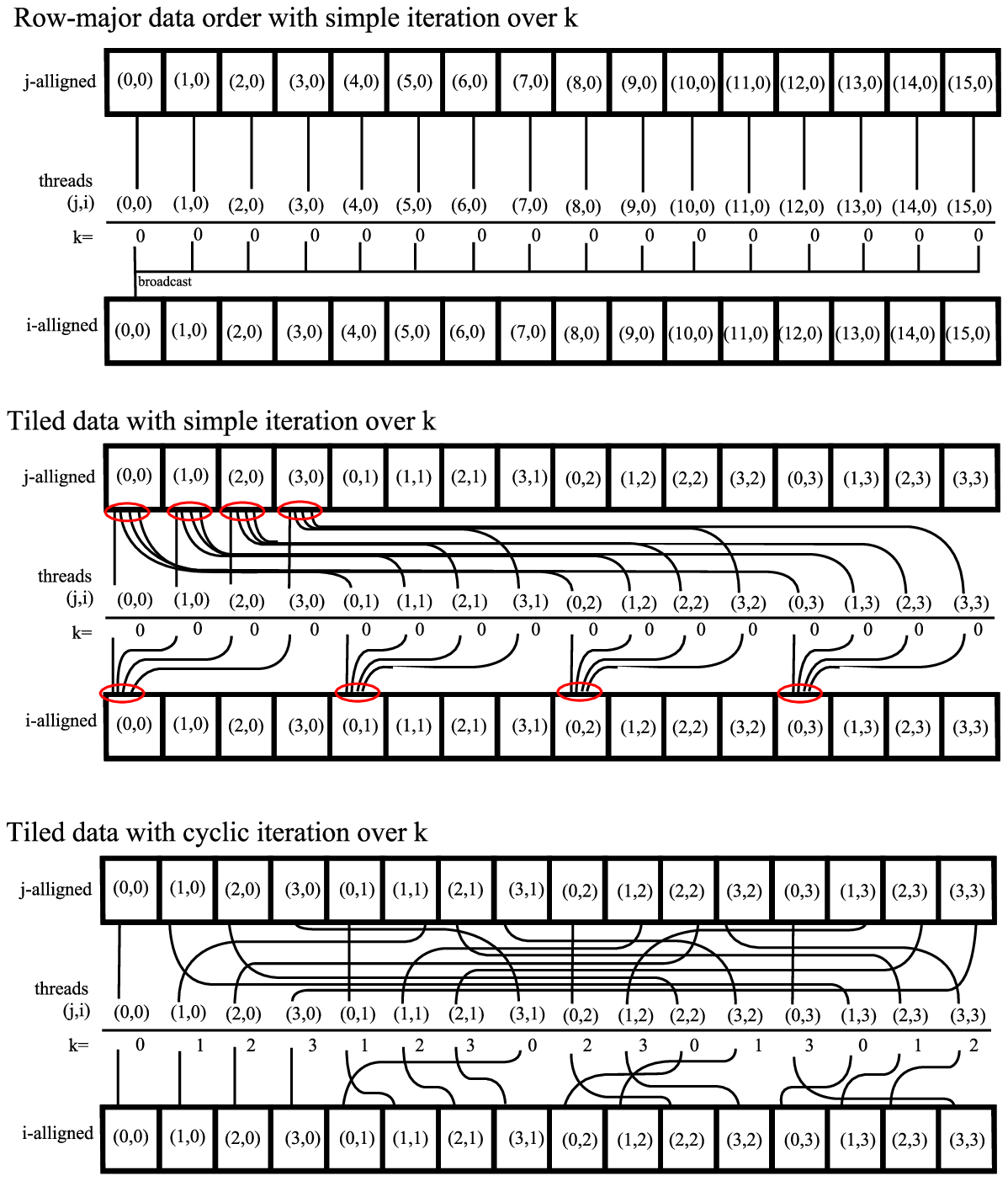}
\caption{Patterns of Shared Memory Access to Singly Dependent Tiles}
\label{SharedMemoryAccess}
\end{figure*}

Figure~\ref{SharedMemoryAccess} shows different patterns of shared memory access that result from using a 4 by 4 tiled memory pattern.
The top figure shows the memory access pattern when the matrix is in row-major order.
In this case, the data elements processed by threads in a half warp are in a row 16 elements long.
Each thread will access an element in the j-aligned tile with the same j-index as it has and k-index determined by the k-iteration it processes.
Each thread will access the same element in the i-aligned tile, since they all have the same i-index.
This results in a conflict free pattern of memory access.
Each element accessed in the j-aligned tile is in a different memory bank since they fall on 16 contiguous words.
The single element in the i-aligned can be broadcast to all of the threads in a single cycle.

The middle figure shows what happens if a 4 by 4 tiled memory pattern is used with the same iteration over k as in the previous case.
Since the threads in a half warp each have one of 4 j-indexes and one of 4 i-indexes, there are four different data elements in each tile that are accessed.
This results in a 4 way data conflict, which uses at least 4 cycles per data access.
Note that the same situation will arise if the data elements updated by threads in a half warp are arranged in a row of 16.
Although the same indexes will be accessed as in the case using a row-major data ordering, these data elements will share data banks.
For example, the data elements in the j-aligned tile having relative indexes $(0,0)$ and $(4,0)$ will be in the same bank.

The bottom figure shows the approach we used to solve this problem.
Instead of iterating over tasks $(i,j)_k$ to $(i,j)_{k+4}$ in order, the task chosen depends on the position of the thread in it's half warp as described above.
This results in conflict free shared memory data access.

\section{Results and Analysis}
\label{sec:Results}
The tests were run on a computer having 1300 MHz \emph{AMD Phenom\texttrademark 9950 Quad-core} processors and  \emph{NVIDIA Tesla C1060} GPUs.
No program used more than 1 processor or GPU.
All programs were compiled as 32-bit applications.

\begin{figure*}
\centering
\includegraphics[width=5in,height=3in]{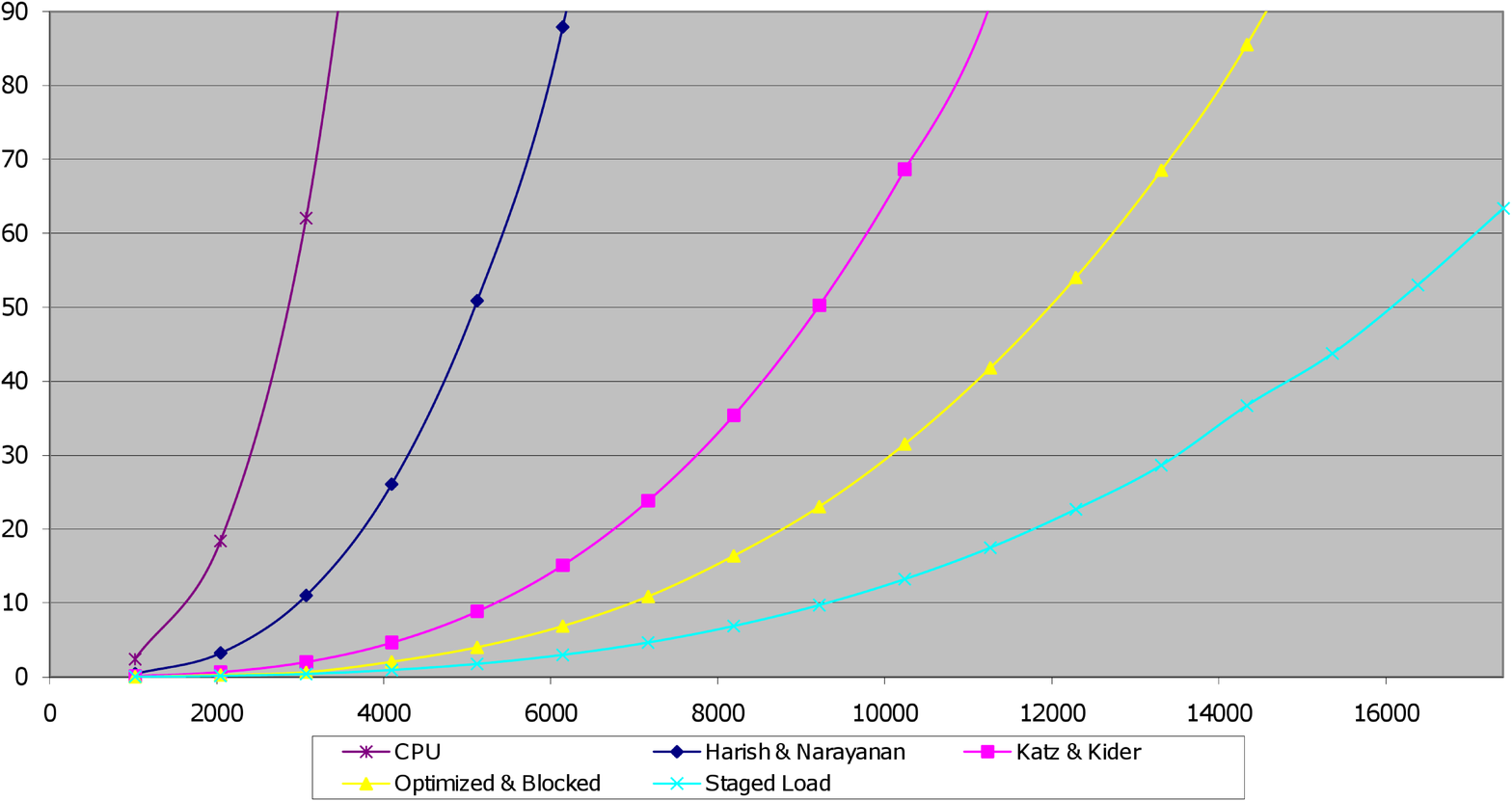}
\caption{Graphs of Results}
\label{GRAPH-Results}
\end{figure*}

Table \ref{TBL-Results} and figure \ref{GRAPH-Results} show the performance of our various implementations on different problem sizes.

The CPU implementation is a basic implementation on the CPU.
The Harish and Narayanan algorithm assigns a single thread for each task performed.
The Katz and Kider algorithm overcomes the bandwidth limitations faced by Harish and Narayanan with blocking.
The blocking size used is 32.
The Optimized and Blocked algorithm is a version of the Katz and Kider algorithm that applies a number of optimizations to reduce the amount of time spent calculating results.
The Staged Load algorithm reduces the latency exposed in the Katz and Kider algorithm by reducing the amount of shared memory used.

\begin{table*}
\label{TBL-Results}
\centering
\caption{Implementation Comparison Times}
\begin{tabular}{ |c||c|c|c|c|c|}
\hline
& \multicolumn{5}{|c|}{\textbf{Times}} \\
\hline
\textbf{Vertices}&\textit{CPU}&\textit{Harish \& Narayanan}&\textit{Katz \& Kider}&\textit{Optimized \& Blocked}&\textit{Staged Load}\\
\hline \hline
\textit{1024} &\texttt{2.405}	&\texttt{0.408}&\texttt{0.108}&\texttt{0.0428}&\texttt{0.0274}\\
\hline
\textit{2048} &\texttt{18.38}	&\texttt{3.212}&\texttt{0.65}	&\texttt{0.282} &\texttt{0.14}  \\
\hline
\textit{3072} &\texttt{62.04}	&\texttt{10.99}&\texttt{2.01}	&\texttt{0.653} &\texttt{0.401} \\
\hline
\textit{4096} &\texttt{145.2}	&\texttt{26.05}&\texttt{4.62}	&\texttt{2.06}	 &\texttt{0.934} \\
\hline
\textit{5120} & &\texttt{50.87}	&\texttt{8.84}	&\texttt{4.02}	&\texttt{1.76}\\
\hline
\textit{6144} & &\texttt{87.9}	&\texttt{15.09}&\texttt{6.89}	&\texttt{2.98}\\
\hline
\textit{7168} & & &\texttt{23.82}	&\texttt{10.9}	&\texttt{4.65} \\
\hline
\textit{8192} & &\texttt{208.6}	&\texttt{35.37}&\texttt{16.39}&\texttt{6.88} \\
\hline
\textit{9216} & & &\texttt{50.24}	&\texttt{23.05}&\texttt{9.71}\\
\hline
\textit{10240}& & &\texttt{68.67}	&\texttt{31.52}&\texttt{13.22}\\
\hline
\textit{11264}& & &\texttt{91.08}	&\texttt{41.82}&\texttt{17.48}\\
\hline
\textit{12288}& & & &\texttt{54.05}	&\texttt{22.67}\\
\hline
\textit{13312}& & & &\texttt{68.56}	&\texttt{28.63}\\
\hline
\textit{14336}& & & &\texttt{85.56}	&\texttt{36.7}\\
\hline
\textit{15360}& & & & &\texttt{43.74} \\
\hline
\textit{16384}& & &\texttt{277.8}	&\texttt{126.9}&\texttt{53.02}\\
\hline
\textit{17408}& & & & &\texttt{63.4}\\
\hline
\end{tabular}
\end{table*}

The Harish and Narayanan implementation is limited by the bandwidth of the global memory bus.
Since each task requires 16 bytes of total traffic over the bus, the bandwidth achieved by this algorithm is 42 GB/sec, somewhat short of the 77 GB/sec bandwidth measured on our device for device-to-device memcpy.

The Katz and Kider implementation is limited by the time spent on calculations.
It can perform $14.9*10^9$ tasks per second.
Based on the advertised speed of 933 GFLOP/sec for a \emph{NVIDIA Tesla C1060}, this is equivalent to using 62.7 FLOPs for each task.
Since each basic task consists of a single add, a minimum, and the accessory operations of locating and loading at least one dependency in shared memory and assigning the result of the calculation to a register, we could reasonably hope to get the actual total processing to under 10 FLOPs per task.

Our staged load implementation performs approximately $73.6*10^{9}$ tasks per second.
If it is limited by bandwidth, it achieves 46 GB/sec, which is less than the 70 GB/sec or so we could reasonably hope for.
If it is limited by the processing speed, it is using the equivalent of 12.7 FLOPs per task.
Since small changes to the code executed on the multiprocessor affect performance, we believe that it is still limited by the processing speed.
In either case, it is possible that further optimizations could improve the performance of this blocked algorithm by a factor of 1.5 or less.

\section{Acknowledgments}
We would like to thank George Purdy for suggesting the topic and Fred Annexstein for his insight and advice.

\bibliographystyle{IEEEtran}
\bibliography{APSP_CUDA}

\end{document}